\documentclass[manuscript,screen]{acmart}
\AtBeginDocument{%
  \providecommand\BibTeX{{%
    \normalfont B\kern-0.5em{\scshape i\kern-0.25em b}\kern-0.8em\TeX}}}

\setcopyright{acmlicensed}
\copyrightyear{2018}
\acmYear{2018}
\acmDOI{XXXXXXX.XXXXXXX}

\acmConference[x '24]{Make sure to enter the correct
  conference title from your rights confirmation emai}{}{USA}
\acmBooktitle{x '24} 
\acmISBN{978-1-4503-XXXX-X/18/06}

\begin{document}


\title{Enabling BLV Developers with LLM-driven Code Debugging}
\author{Clark Saben}
\email{csaben@mail.umw.edu}

\author{Prashant Chandrasekar}
\email{pchandra@umw.edu}

\affiliation{%
  \institution{University of Mary Washington}
  \city{Fredericksburg}
  \state{Virginia}
  \country{USA}
  \postcode{22401}
}

\renewcommand{\shortauthors}{}

\begin{abstract}

BLVRUN is a command line shell script designed to offer developers within the BLV community a succinct and insightful overview of traceback errors. Its primary function involves parsing errors and utilizing a refined large language model to generate informative error summaries. In terms of performance, our model rivals that of well-known models like ChatGPT or AI-chatbot plug-ins tailored for specific Integrated Development Environments (IDEs). Importantly, BLV users can seamlessly integrate this tool into their existing development workflows, eliminating the need for any modifications or adaptations to facilitate debugging tasks.

\end{abstract}


\maketitle

\section{Introduction}
Code debugging is on one of the five main programming tasks or challenges that developers undertake while building programs or softwares \cite{mountAbou}. 
Whether one is new to coding, or an experienced coder, one comes across the situation where their code, or the code they're working with, causes an error.
We are specifically describing a situation where executing the code produces an explicit error. 
Our research does not address the scenario of unintended behavior where no error is produced on the screen.
These ``traceback'' errors are unstructured text that is often verbose and difficult to comprehend. 
While the intention is to communicate as much information as possible, the source of the error, and the understanding of it, can only be discovered by putting together information from interpreting one or two key sentences. 
Developers of the blind and low vision (BLV) community are forced to process the lengthy error trace sequentially.

This process can be time-consuming as it extends the time and effort towards debugging and, thereby, writing correct code.   
BLVRUN is a shell script that runs in the background of any command prompt on any operating system.
Once installed, it takes in error trace and summarizes it using a machine learning model that has been fine-tuned on the widely used Python bug dataset, PyTraceBugs \cite{pytracebugs}.
The BLV programmer (user) is then presented only with a concise and insightful summary of the error trace. 
While code summarization or code description through machine learning is not new, our solution focuses on maintaining the current workflow of BLV programmers who use text buffers, printf-styled debugging methods, and command line.
By using BLVRUN, BLV programmers do not need to make any change to their current workflow.
Neither would they have to worry about committing to a ``tool'' and keep up with the tool or plug-in's documentation and development lifecycle.
From their point-of-view, when an buggy program is executed, a summarized error is produced.
This will greatly reduce the time it takes for BLV programmers to assess an error, thereby reducing, if not removing, the frustrations that BLV programmers would otherwise experience. 
For BLV programmers in the early stages of learning the art, this would be hugely impactful.  
\section{Related Work}
This research effort is informed by two specific (sub-)areas of literature. The first area focuses on code debugging challenges. 
Code Debugging is listed as one of the primary programming tasks and/or challenges that programmers undergo in their development process \cite{aboubakar}.
Debuggers are primarily used for either: (a) analyzing runtime behavior, or (b) finding logical errors in code.
Our research specifically addresses BLV developers who deal with (b).
Debugging or, more specifically, the process of finding errors in code is heavily supported in IDEs. 
There are numerous reasons why current debugging tools or plugins are challenging to use for BLV developers.
Firstly, use of debugging tools or plugins within IDEs require understanding of the spatial layout of IDEs. 
The functionality of the debugging tool is built without consideration of screen reader interpretation, leading to incompatibility between the two \cite{stefik, albusays, potluri, bakercat}.
Visual aids such as Syntax Highlighting or Syntax Error cues like squiggles, that assist in code debugging and possibly navigation, is challenging for programmers with visual impairments, especially when compared against sighted programmers \cite{pandey, potluri}.
This results in BLV developers leaning on text editors instead of IDEs, employing ``printf'' debugging instead, and editing code using text buffers \cite{mealin, albusays}. 
These ``workarounds'' imply the requirement or use of the command-line interface in conjunction \cite{harini2021, pandey2021}.
This brings us to our second context or area of our research. 
Harini et. al highlight the lack of understanding of accessibility of CLIs \cite{harini2021}.
One of the main findings of their work was the inaccessibility of scrolling a terminal with screen readers. 
Scrolling is extremely pertinent in the context of debugging.
Not even considering the outcome of ``printf debugging'', scrolling through traceback errors is a challenge.
We calculated the some metrics on a large Python software defect dataset.
The median number of sentences per error, among 3864 errors, was around 26.
The median number of words per error was 76. 
Which is why they recommended that (a) long output of unstructured text is converted into an accessible format, and (b) error messages be easier to comprehend.

The primary intent of this research is to aid BLV developers in their debugging efforts by improving the accessibility of CLI output. 
In the following sections we describe the design of our solution and report on its performance. 

\section{Solution}

\subsection{Approach}
BLVRUN, our innovative CLI application, is designed to simplify the debugging process by providing concise summaries of traceback errors. It is built on a two-pronged approach:

\begin{figure}[!htp]
  \centering
  \includegraphics[width=0.90\linewidth]{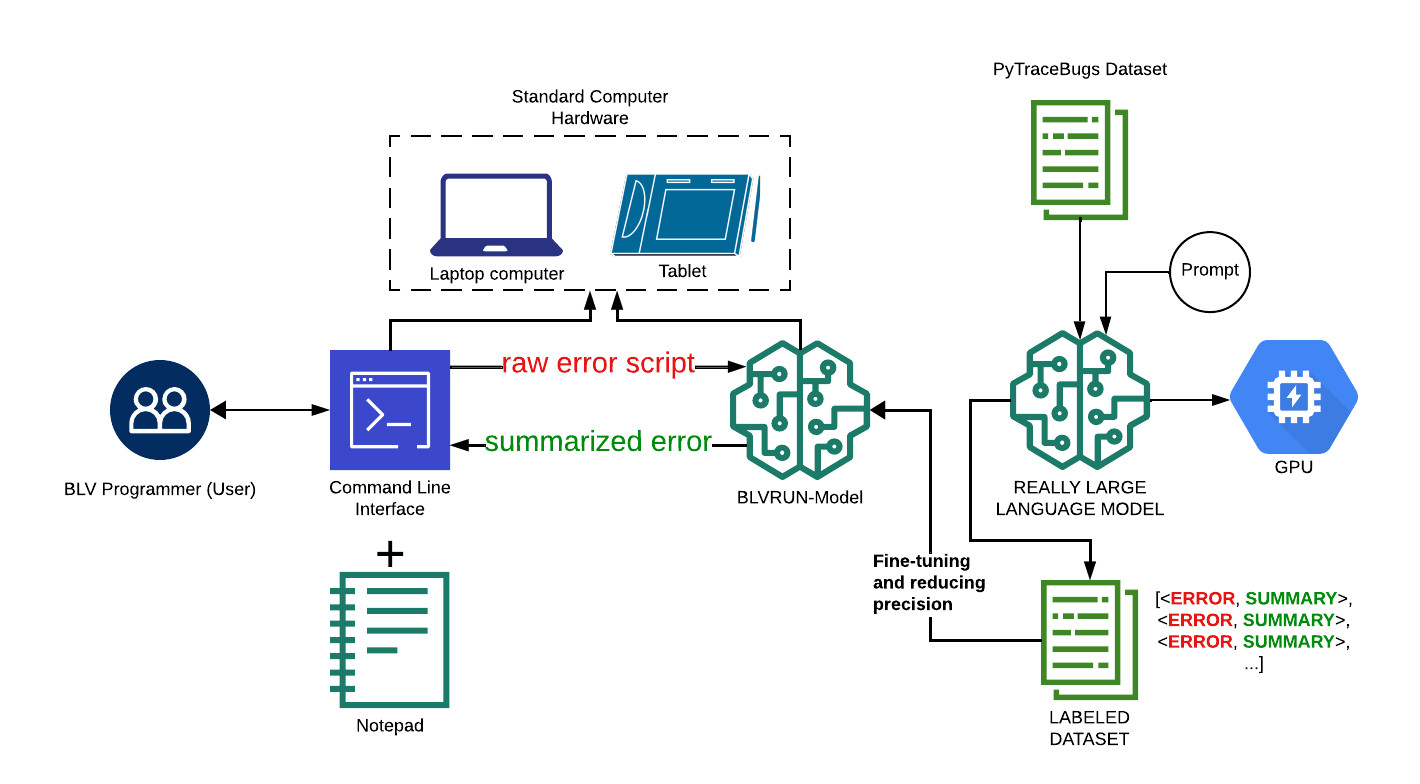}
  \caption{Architecture and Development Components of BLVRUN. Starting from the left, a BLV programmer, who using CLI and text buffers executes their Python code. When an error is produced, BLVRUN's script captures the verbose and unstructured text and only presents the user with a concise and accurate description of the error. This is possible because BLVRUN's model is fine-tuned using a dataset we created from PyTraceBugs. Finally, BLVRUN is optimized to run on any machine, thereby not requiring BLV programmers to depends on IDEs and/or switch contexts with ChatGPT-like solutions.}
  \Description{Finetuning Diagram}
  \label{fig:approach}
\end{figure}

\begin{enumerate}
    \item A Rust-written shell application that operates in the background, monitoring the output of Python code executions. This script is adept at capturing the often unstructured and verbose text generated from traceback errors.
    \item A fine-tuned 7 billion parameter CodeLlama model \cite{cllamaroziere2023code}. This model is specifically trained on traceback data, ensuring that even with reduced precision, it maintains robust performance on traceback summarization, even if its efficiency in other tasks diminishes.
\end{enumerate}

LLMs are typically slow in generating text on standard consumer-grade hardware. 
To counter this, we have optimized BLVRUN by reducing the precision of the model's parameters. 
While this makes text generation faster compared to using full precision, it's still not rapid enough for an optimal user experience. 
To further address this, BLVRUN is configured to load the precision-reduced model upon startup. 
Consequently, every time an error is emitted in the terminal, BLVRUN swiftly captures the traceback information and consults the model for a summary.

This setup ensures that users do not need to adopt any new practices, gain additional knowledge, switch contexts, or acquire new hardware to better understand their code errors. 
In the subsequent (sub-)sections we delve into the details of BLVRUN's development. 
This is also shown in Figure \ref{fig:approach}.

\textbf{
\subsection{Dataset}
}
The dataset enhancing BLVRUN's model performance is PyTraceBugs \cite{pytracebugs}.
This dataset includes training and evaluation data with 14,118 and 56 errors, respectively.
The PyTraceBugs dataset offers a broad spectrum of traceback types.
This is beneficial for model generalization and, therefore, user support.
There are 555 unique error types across the dataset.
However, the occurrence per error type is extremely sparse.
Additionaly, we filter both training and test sets for the keyword "Traceback," focusing on this aspect in our model.
We wanted to ensure that our evaluation of our model was rigorous. 
Therefore, we only fine-tuned and tested our model on certain categories of errors that were most commonly occurring in both the training set and the test set. 
These traceback errors include; TypeError, ValueError, AttributeError, IndexError, NameError, RuntimeError, and KeyError. Specifics about these errors in the dataset can be found at \cite{pytracebugs}.

\textbf{
\subsection{Fine-Tuning}
}

Fine-tuning is a technique where we take a model that has already been trained on a general task and then continue training it on data specific to a particular domain.
This process often results in the model performing better on the new, specific task.
In the case of BLVRUN, the base model is provided by Meta and is known as CodeLlama \cite{cllamaroziere2023code}. Originally trained on Python code, CodeLlama serves as an advanced development assistant, generating Python code based on the context provided.

CodeLlama operates by taking the given code and predicting the next sequence of tokens, which are essentially bits of Python code.
Meta has released several versions of the CodeLlama model, each differing in the amount of computational power and memory required.
This variance is due to the number of parameters in each model.
Parameters, the values in each layer of the neural network, are crucial in the process of output generation. The CodeLlama models come in different sizes, namely 7 billion, 13 billion, and 34 billion parameters, with larger models typically showing better performance.

The fine-tuning process itself is done using QLoRA \cite{dettmers2023qlora}. QLoRA is an advanced technique designed to make training large AI models, like the one in BLVRUN, more manageable on regular computers. Training such extensive models typically requires substantial computing power, but QLoRA reduces this need by cleverly minimizing memory usage.
At its core, QLoRA specializes in fine-tuning large models (with billions of parameters) while still maintaining high-quality performance. Think of it like precisely adjusting a complex machine to improve its efficiency. In the context of AI, it involves refining certain model components to enhance its task-specific effectiveness.
A standout feature of QLoRA is its use of a novel data type called 4-bit NormalFloat (NF4). This data type is theoretically optimal for handling weights in the model that follow a normal distribution, a common scenario in AI model data. Essentially, it’s akin to finding an incredibly efficient way of packing data into a smaller space without losing the essence of the information.
These high-performance models offer a progressive approach to handling complex tasks.
For instance, we use the larger parameter model to process a subset of the errors in the training set to generated our training data of ``<error, summary>'' pairs. 

To build BLVRUN, we use the 7 billion parameter CodeLlama as the base model.
We fine-tune the base model  using the dataset generated from PyTraceBugs, in combination with the 13 billion parameter CodeLlama model (as mentioned previously).
This approach helps BLVRUN to better understand and handle Python traceback errors, which are critical in debugging and development processes

\textbf{
\subsection{Reducing precision}
}
A common practice in deploying Large Language Models (LLMs) involves reducing the model's precision beforehand. 
This process is known as quantization. 
Reducing precision refers to storing fewer digits after the decimal point for each parameter in the model. 
The lower the precision, the smaller the model, thereby increasing the speed at which it can generate summaries.
However, this approach entails a performance tradeoff. 
As the precision decreases, so does the model's representation quality, which can affect its effectiveness.

By concentrating on a specific task, the model is expected to maintain acceptable performance levels in that area, despite a potential drop in its overall capabilities.

The process begins by applying QLoRA. 
After the QLoRA fine-tuning, a specialized component called a LoRA adapter is produced. 
This adapter contains the adjustments made to the original model, enabling it to perform well on specific tasks despite the reduced precision. 
The LoRA adapter is then converted to match the format of the original model, ensuring compatibility.
Subsequently, the base model and the LoRA adapter are merged. 
This combined model is further quantized to what is termed Q2K, where on average, each parameter is represented by approximately 2 bits. 
This extreme level of quantization greatly reduces the model's size from 12.55GB to just 2.83GB, making it more manageable for deployment in applications like BLVRUN.

As described further in the Evaluation section, the Q2K quantization achieves significant performance gains, especially when compared to a model that is simply quantized to 2 bits without the benefit of our QLoRA fine-tuning.

\subsection{User interface}

\begin{figure}[!htp]
  \centering
  \includegraphics[width=\linewidth]{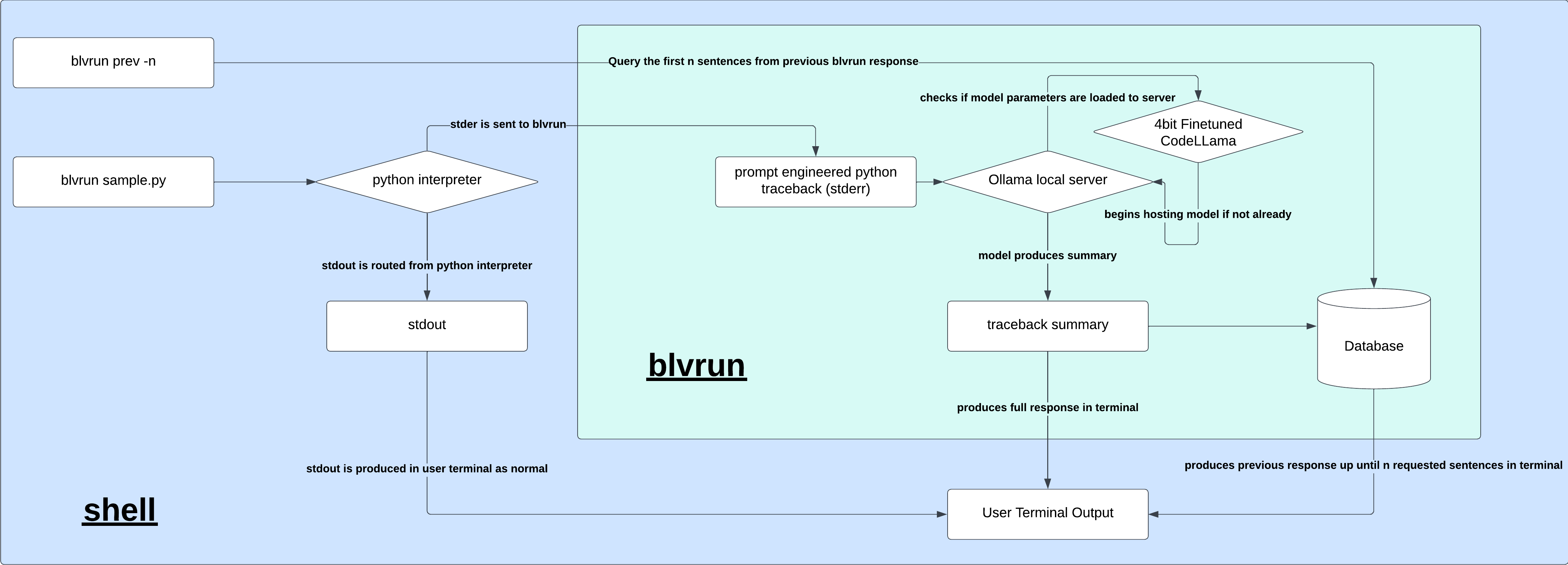}
  \caption{Information Flow within BLVRUN. When blvrun sample.py is executed in the shell, the prompt is sent to our model that is hosted on a Ollama server. Our model produces a traceback summary that is sent back to the terminal and saved in a database. BLV programmers can see previously generated summaries using the blvrun prev -n command. }
  \Description{UI Diagram}
  \label{fig:blvrunarch}
\end{figure}

\begin{figure}[!htp]
  \centering
  \includegraphics[width=0.90\linewidth]{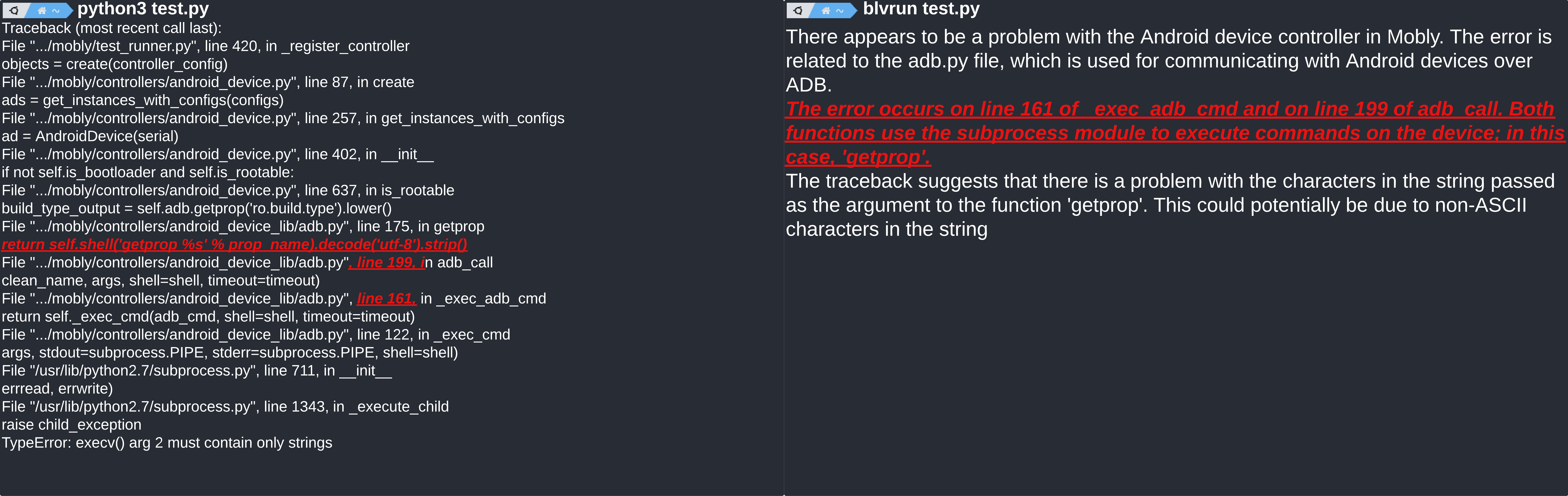}
  \caption{Example of the usefulness of BLVRUN. On the left one can see the unstructured, and verbose, output printed to BLV programmers (without the assistance of BLVRUN). On the right, we see the summary produced by BLVRUN. Within the summary, we have highlighted the key takeaway of the error, which BLVRUN presents it in a concise and, therefore, consumable manner.}
    \label{fig:userinteraction2}
  \Description{big traceback summary diagram}
\end{figure}

The user interface of BLVRUN is command-line driven, designed to assist users in quickly understanding and acting on Python traceback errors. 
When a Python script encounters an error, as seen in the right of Figure \ref{fig:userinteraction2} executing \verb|blvrun sample.py| directs the traceback to BLVRUN, which then provides a clear summary of the error in the terminal.
Where BLVRUN really improves quality of life while coding for BLV users can be seen when compared to the error shown on the left of Figure  \ref{fig:userinteraction2}. 
A large traceback will often verbosely list line numbers from dependency files where each error occurs while not necessarily making it clear where in the user's code the issue is on a glance. 
BLVRUN provides an excellent workaround with allowing the model to read and summarize where the error likely occurs.

BLVRUN allows users to revisit the last generated summary with a command such as \verb|blvrun prev -n|, enabling them to retrieve the 'n' most recent sentences from the summary for further review. 
This feature is particularly useful for users who need to recall or further examine the details of the last error without re-running the script. 
The database supporting this feature is optimized to hold just the previous response, maintaining system efficiency by avoiding storage of extensive historical data.

The focus of BLVRUN's UI is on functionality and ease of use, offering essential commands without overwhelming the user with unnecessary options or configurations. 
This ensures that the user can remain focused on their primary task-coding, while BLVRUN handles error summarization efficiently in the background. 

\section{Evaluation}

\begin{figure}[!htp]
  \centering
  \includegraphics[width=\linewidth]{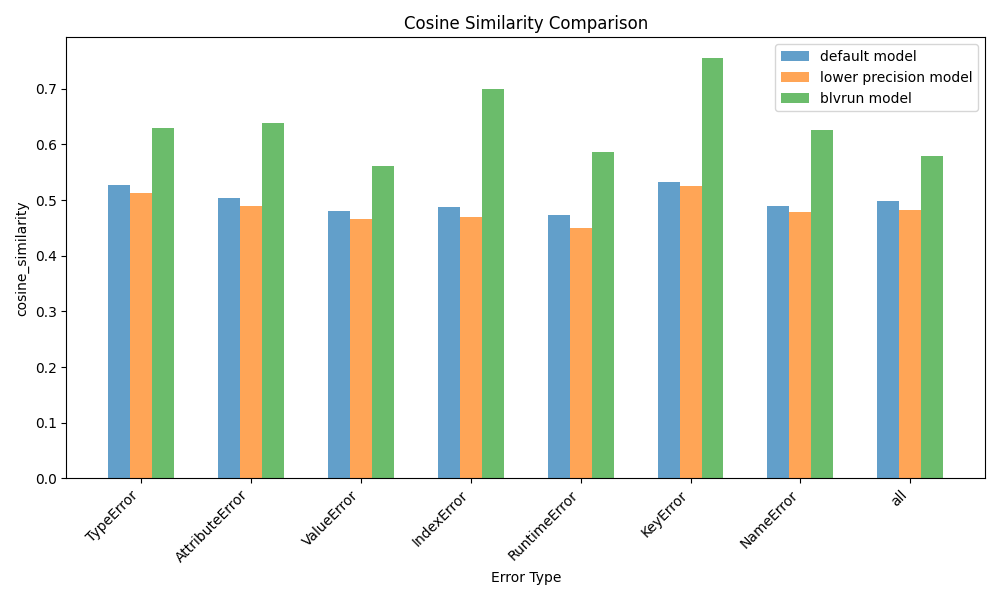}
  \caption{Cosine similarity scores of summaries generated by (1) base model (with no fine-tuning or lowered precision), (2) base model (with lowered precision), (3) BLVRUN's fine-tuned and optimized model compared against ``gold standard''}
    \label{fig:cosinesimilarity}
\end{figure}

\begin{figure}[!htp]
  \centering
  \includegraphics[width=\linewidth]{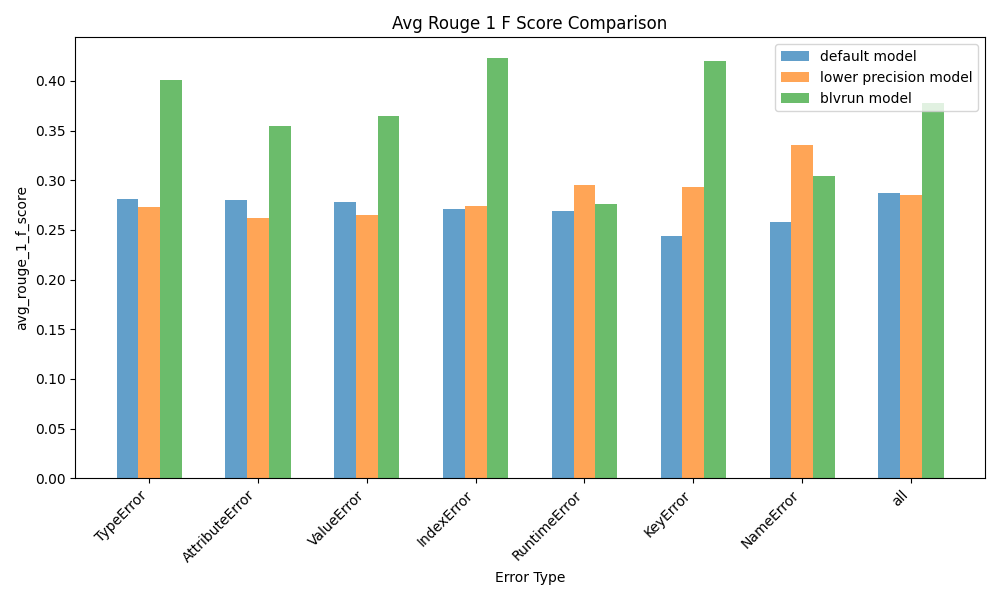}
  \caption{ROUGE-1 f-scores of summaries generated by (1) base model (with no fine-tuning or lowered precision), (2) base model (with lowered precision), (3) BLVRUN's fine-tuned and optimized model compared against ``gold standard''}
    \label{fig:rouge1}
\end{figure}

We designed BLVRUN to run on any computer hardware in the background of the terminal without disrupting a BLV user's workflow.
BLVRUN is only helpful if it produces accurate summaries of long traceback errors. 
To evaluate the accuracy, we compared the output produced from BLVRUN against Meta's largest model (13 billion parameter model). 
PyTrackback's open dataset also provides a test set to be used for evaluation. 
Similar to the fine-tuning process, we employed Meta's model to generate a labeled dataset of ``<error, summary>'' pairs  from the test set of PyTraceback. 
As a result, we compared BLVRUN's summaries against this ``gold-standard'' for 23 number of errors. 
We picked a subset of the errors based on the number of data points available for each error.
We used cosine similarity and ROUGE-1 to compare the two summaries.
From Figures \ref{fig:cosinesimilarity} and \ref{fig:rouge1}, we see that our model achieve a fairly high similarity, and high ROUGE-1 scores.
Most importantly, we can see the improvement achieved in the model performance when compared against the baseline non-fine-tuned model and non-optimized model. 
We find that BLVRUN does the best job in producing a shorter version of a lengthy traceback error by excluding the ``noisy'' text and highlighting the key insights.
In the next section we describe our plans to expand on our model capabilities and our planned user study.

\section{Planned Future Work}
Our current and near-future efforts are two-pronged:
(1) Widen the support for prompt message understanding: 
Currently, our model is fine-tuned to support error messages that were frequently represented in the error database.
Our immediate next step is to widen the support for messages, error or otherwise, by generating a synthetic dataset to augment the low-incidence report of these messages. 
(2) Gain insight and feedback to inform future design of BLVRUN: 
We are in the process of getting IRB approved for our user study. 
We plan to recruit BLV participants who primarily use text editors and command line terminals in their current development workflow.
Participants in our control group will be asked to perform tasks that involve interaction with the terminal without support of BLVRUN.
Participants in our test group will perform the same tasks using the aid of BLVRUN. 
We will measure task performance, usability of the BLVRUN interface, and self-reported impact on participants (such as stress, fatigue, frustration, etc.). 
Insights from this study will further inform the development of future AI-models and design of the interface.
\section{Contributions}
AI- or LLMs-centric models are reliable enough that they are increasingly being included as part of the interface for programming IDEs. 
Our effort has demonstrated the feasibility of harnessing that potential and reliability into a command line interface that is considered, through support from many studies, as being more accessible as compared to IDEs.
More importantly, our solution works on a local CPU, thereby not requiring BLV users to switch contexts or change their current programming ``workflow.'' 
BLVRUN is currently fine-tuned and engineered to support more frequently occurring errors.
This is done to support debugging-related efforts through the command line. 
We are refining our models to increase the scope of different types of program outputs that BLVRUN can support.
In our user study, we plan to examine the the long-term impact of BLVRUN on the development process of the BLV programmers.
Any insights will naturally inform the design of our model and future work on its interface.

\bibliographystyle{ACM-Reference-Format}
\bibliography{sample-base}

\end{document}